# A TIGHTER CONSTRAINT ON POST-NEWTONIAN GRAVITY USING MILLISECOND PULSARS

J. F. Bell

*Mount Stromlo and Siding Spring Observatories, Institute of Advanced Studies,*
*Australian National University, Private Bag, Weston Creek, ACT 2611, Australia*
*email: bell@mso.anu.edu.au*

## ABSTRACT

Some theories of gravity predict the existence of preferred-frame effects and violations of conservation of energy and momentum. General relativity predicts no such effects. In the parameterised post-Newtonian (PPN) formalism, the parameter, $\alpha_3 \equiv 0$ if these effects do not exist. The period derivatives ($\dot{P}$) of millisecond pulsars (MSPs) are used to more tightly constrain these effects by showing that $|\alpha_3| < 5 \times 10^{-16}$.

*Subject headings:* gravitation, relativity, stars: pulsars: general

The PPN formalism is a very powerful tool for analysing gravitation theory and experiment. It provides a set of parameters (the PPN parameters) which take different values in different theories and can be related to measurable quantities, forming a basis for comparison of theory and experiment. See Will (1993) for a summary of the PPN formalism, PPN parameters and the values they take in various theories of gravity. The PPN parameter $\alpha_3$ is of particular interest since in addition to being sensitive to violations of conservation of energy and momentum, it is sensitive to preferred-frame effects. That is, it measures the extent to which the motion of a gravitating system through the mean rest frame of the local universe can produce local gravitational effects. Theories (including general relativity) that have no preferred-frame effects and conserve energy and momentum have $\alpha_3 \equiv 0$.

A consequence of $\alpha_3$ being nonzero is a "self-acceleration" of a spinning body as it moves through space. The direction of the acceleration is perpendicular to its spin axis and velocity vector. For a population of radio pulsars the acceleration will be randomly oriented, since the spin axes of pulsars are randomly oriented. An acceleration of a pulsar along the line-of-sight to the pulsar causes a change in the observed $\dot{P}$, analogous to the line-of-sight velocity of a pulsar causing a Doppler shift in its observed rotation period. Hence the change in $\dot{P}$ due to a self-acceleration is $\Delta \dot{P} = P \hat{\mathbf{n}} \cdot \mathbf{a}_{self}$, where $P$ is the rotation period of the pulsar, $\mathbf{a}_{self}$ is the self acceleration and $\hat{\mathbf{n}}$ is the unit vector along the line-of-sight to the pulsar (Will 1993). Since $\mathbf{a}_{self} \propto P^{-1}$, the resulting change in $\dot{P}$ is independent of pulse period. Hence, if such self accelerations exist and are large, the distribution of period derivatives will be broadened towards a distribution with median $\dot{P}$ of zero. This is something we can readily test with a population of pulsars.

This test has been evaluated previously by Will (1992,1993), who considered the period derivatives of the main population of pulsars and obtained $|\alpha_3| < 2 \times 10^{-10}$. Since the MSPs are a separate class of objects from the ordinary pulsars, we can consider the period derivatives of MSPs alone. This will provide a much tighter constraint, as the period derivatives of MSPs are typically 5–6 orders or magnitude smaller than those of ordinary pulsars. The sample of MSPs we selected includes all known MSPs with measured period derivatives, having $P < 20$ ms in the Galaxy (Taylor, Manchester & Lyne 1993, Johnston et al. 1993, Lorimer et al. 1995, Bailes et al. 1994, Camilo, Foster & Wolszczan 1994, Camilo, Nice & Taylor 1993). Those MSPs in globular clusters are excluded, as negative period derivatives for pulsars in globular clusters are known to be caused by acceleration of the pulsar in the cluster potential. Period derivatives of the 18 Galactic MSPs remaining in the sample, may similarly be affected by their acceleration toward the Galactic disk and the acceleration due to differential rotation (Damour & Taylor 1991). While these effects are typically an order of magnitude smaller than the measured period derivatives, they are taken into account in the analysis.

A more important source of the corruption of period derivatives is the apparent line-of-sight acceleration due to the proper motion of pulsars (Shklovskii 1970, Camilo, Thorsett & Kulkarni 1994). The change in $\dot{P}$ due to this effect is always positive and is given by $\Delta \dot{P} = 1.1 \times 10^{-18} v^2 P/cd$, where $v$ is the tangential velocity in units of 100 kms$^{-1}$ and $d$ is the distance in kpc. For PSR J0437-4715 this effect contributes 2/3 of the measured period derivative (Bell et al. 1995). This was corrected for prior to analysing the period derivatives. In the sample of MSPs, 12 have measured velocities (Nice & Taylor 1995, Bell et al. 1995, Nicastro & Johnston 1995, Camilo, Thorsett & Kulkarni 1994) and their median velocity of 69 kms$^{-1}$ was used for the other 6 with undetermined velocities.

Figure 1 shows the cumulative distribution function (CDF) of the corrected period derivatives for the sample of 18 MSPs. Clearly a population with median $\dot{P} = 0$ is inconsistent with what is observed. The rms $\dot{P}$ for the observed distribution may be due to both the rms of the intrinsic distribution and contributions from self accelerations that may exist. Hence, we can use the observed rms $\dot{P}$ of $2.5 \times 10^{-20}$, to set a limit on $\alpha_3$. Following the evaluation of $\hat{\mathbf{n}} \cdot \mathbf{a}_{self}$ by Will (1993), we have $\dot{P} \simeq 5 \times 10^{-5} |\alpha_3|$, which gives $|\alpha_3| < 5 \times 10^{-16}$, a limit 5.6 orders of magnitude tighter than the previous best. The major source of uncertainty here is the poorly determined corruption of $\dot{P}$ due to the proper motion of MSPs, resulting from distances that are accurate to only 30%. This is particularly important since the effect increases the observed period derivative. However, for this effect to be responsible for increasing the observed median $\dot{P}$ from zero to $1.5 \times 10^{-20}$, would require the MSP population to have a median tangential velocity of 170 kms$^{-1}$. Since the observed median tangential velocity for MSPs is 69 kms$^{-1}$, this is unlikely. This new bound on $\alpha_3$ together with bounds on the other 9 PPN parameters (Will 1993) confirm all the predictions of general relativity.

We thank Matthew Bailes, Dick Manchester, Prasenjit Saha and the referee, Clifford Will for helpful suggestions. JFB received support from an Australian Postgraduate Research Award and the Australia Telescope National Facility student program.

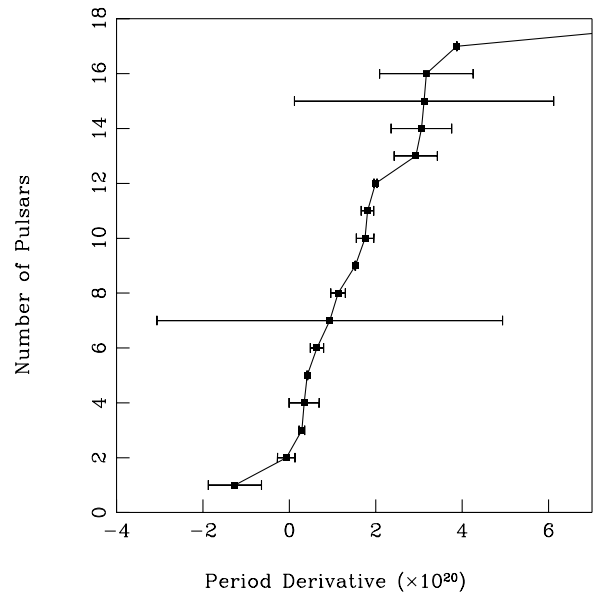

Fig. 1.— The cumulative distribution function of the period derivatives of a sample of 18 millisecond pulsars. Error bars show the uncertainty in individual measurements after correction for acceleration effects.